\documentclass[english,aps,showpacs,prb,twocolumn,floatfix,superscriptaddress,a4paper]{revtex4}
\usepackage{amsmath}
\usepackage{graphicx}
\usepackage{amssymb}
\usepackage{color}
\usepackage{hyperref}
\usepackage{natbib}
\usepackage{dsfont}
\usepackage{units}
\usepackage{placeins}
\usepackage{bm}
\usepackage{dcolumn}

\makeatletter
\usepackage{babel}
\makeatother

\begin{document}
\onlinecite{} 
\title{Effect of strain on hyperfine-induced hole-spin decoherence in quantum dots}
\date{\today}
\author{Franziska Maier}
\affiliation{Department of Physics, University of Basel, Klingelbergstrasse 82, CH-4056 Basel, Switzerland}
\author{Daniel Loss}
\affiliation{Department of Physics, University of Basel, Klingelbergstrasse 82, CH-4056 Basel, Switzerland}
\begin{abstract}
We theoretically consider the effect of strain on the spin dynamics of a single heavy-hole (HH) confined to a self-assembled quantum dot and interacting with the surrounding nuclei via hyperfine interaction. 
Confinement and strain hybridize the HH states, which show an exponential decay for a narrowed nuclear spin bath. 
For different strain configurations within the dot, the dependence of the spin decoherence time $T_2$ on external parameters is shifted and the non-monotonic dependence of the peak is altered. 
Application of external strain yields considerable shifts in the dependence of $T_2$ on external parameters. 
We find that external strain affects mostly the effective hyperfine coupling strength of the conduction band (CB), indicating that the CB admixture of the hybridized HH states plays a crucial role in the sensitivity of $T_2$ on strain. 
\end{abstract}
\maketitle

%
\section{Introduction}
During the last years, heavy-hole (HH) spins have attracted much interest in the field of spintronics and spin-based quantum computing. 
This is because, compared to the timescales set by the electron spin, very long hole spin relaxation times $T_1 \approx1 \mathrm{ms}$ have been predicted \cite{Bulaev2005,Trif2009} and confirmed experimentally \cite{Heiss2007,Gerardot2008}. 
Furthermore, ensemble hole spin coherence times $T_2^*>100\mathrm{ns}$ have been measured \cite{Brunner2009}. 
Alongside, the essential ingredients for processing quantum information successfully, hole spin initialization \cite{Gerardot2008,Eble2010} and coherent control of single hole spins  \cite{Greilich2011,DeGreve2011,Godden2012}, have been shown in quantum dots. 
Additionally, methods applicable to overcome decoherence, e.g.\ by preparing the nuclear spin bath in a narrowed state \cite{Coish2004,Klauser2006,Stepanenko2006,Greilich2007,Reilly2008,Vink2009,Bluhm2010}, have been introduced.
The prolonged timescales regarding decoherence are attributed to the Ising-like hyperfine coupling of holes \cite{Fischer2008} due to the $p$-wave symmetry of the Bloch states in the valence band (VB). 
Recently, the hyperfine interaction strength of holes was predicted to be approx.\ 10\% of the interaction strength of electrons \cite{Fischer2008}.
This was confirmed in experiments carried out in self-assembled InAs quantum dots \cite{Fallahi2010,Chekhovich2011}.
The associated  hole spin decoherence time $T_2$ was shown to depend on external parameters in a non-monotonic fashion \cite{Fischer2010}.
Due to lattice mismatch, the strain profiles of InAs/GaAs dots show a compression in the lateral plane and a stretching in the vertical direction \cite{Tadic2002}. 
The associated strain fields are of considerable strength and may strongly affect the band hybridization in the dot and hence the spin decoherence. 
For the light hole (LH) and HH band, the effect of confinement and anisotropic lateral strain on band mixing and on the interaction with a Gaussian nuclear field distribution via dipole-dipole hyperfine interaction has been considered in Refs.~\onlinecite{Eble2009,Testelin2009}. \\
%
In the present work, we examine the effect of realistic strain distributions on the spin decoherence time $T_2$ of a single HH spin confined to a self-assembled InAs quantum dot interacting with a narrowed nuclear spin bath via hyperfine interaction. 
We follow the procedure outlined in Ref.~\onlinecite{Fischer2010} with emphasis on the new features coming from strain. 
The emerging band hybridization is strain dependent and shows considerable admixtures of the lowest conduction band (CB), and the LH and the split-off (SO) band of the VB.
An effective hyperfine Hamiltonian is derived from the hybridized states being, for realistic strain configurations, predominantly of Ising form with small hole-nuclear-spin flip-flop terms which cause exponential spin decoherence. 
We study the effect of various internal strain configurations and of applied external strain on the decoherence rate $1/T_2$ and its dependence on external parameters. 
Applying external strain up to the breaking limit of the sample affects the effective hyperfine coupling of the CB admixture much more than the coupling of the LH admixture. 
In contrast to this, the changes in the Ising like HH coupling are negligible.
This indicates the significance of the CB admixture regarding the changes of $1/T_2$ due to strain. \\
%
The outline for the paper is as follows. 
In Sec.~\ref{sec:HHinstraindot} we introduce the 8$\times$8 $\mathbf{k}\cdot\mathbf{p}$ Hamiltonian describing states confined to a strained quantum dot and calculate the hybridized eigenstates of the HH subsystem. 
We find an effective Hamiltonian which describes the hyperfine interaction of the hybridized HH spin states with the surrounding nuclei in Sec.~\ref{sec:hyperfine}.
In Sec.~\ref{sec:timeevolution} we derive the dynamics of the transverse HH pseudospin states and examine the effect of strain on the decoherence time $T_2$ and on the hyperfine coupling constants. 
A summary can be found in Sec.~\ref{sec:summary}. 
Technical details are deferred to the Appendix. 
\section{Heavy hole states in strained quantum dots\label{sec:HHinstraindot}}
%
%
We use the 8$\times$8 $\mathbf{k}\cdot\mathbf{p}$ Kane Hamiltonian $H_{\text{K}}$ which describes the states of bulk zincblende semiconductors in the lowest CB and in the HH, the LH and the SO band of the VB \cite{Kane1957,Winkler2003}.
We assume a  flat, cylindric dot geometry which is taken into account by choosing harmonic confinement $V_{\text{conf}}$ with lateral and vertical confinement lengths $L$ and $a$, respectively, satisfying $L\gg a$. For detailed expressions of $H_{\text{K}}$ and $V_{\text{conf}}$ see the Appendix. Strain is added perturbatively to the system by employing an 8$\times$8 strain Hamiltonian $H_{\bm{\varepsilon}}$ \cite{Trebin1979,Winkler2003} of the form
%
%
\begin{equation}
  H_{\bm{\varepsilon}} = \begin{pmatrix}
    S_{11}& S_{1} & S_{2} & S_{3}\\
    S_{1}^{\dagger} & S_{22} & S_{4} & S_{5}\\
    S_{2}^{\dagger} & S_{4}^{\dagger} &S_{33} & S_{6}\\
    S_{3}^{\dagger} & S_{5}^{\dagger} &S_{6}^{\dagger} & S_{44}
  \end{pmatrix},
\end{equation}
where the relevant block matrix elements are
\begin{equation}
\begin{array}{ll}
  S_1 = \begin{pmatrix}E^*&0\\0&-E\end{pmatrix}, &
  S_{22} = \begin{pmatrix}F+G&0\\0&F-G\end{pmatrix}, \\
  &\\[-3mm]
  S_4 =\begin{pmatrix}I&J^*\\J&I^*\end{pmatrix},&
  S_5 = \frac{1}{\sqrt{2}}\begin{pmatrix}-I & -2J^* \\ 2J & I^*\end{pmatrix}. 
\end{array}
\end{equation}
The entries read $E=\sqrt{2}P[k_x \varepsilon_{xx} -i k_y \varepsilon_{yy}]$, $F = D_d\, \mathrm{Tr}\bm{\varepsilon}-1/3\, D_u(\varepsilon_{xx}+\varepsilon_{yy}-2\varepsilon_{zz})$, $G = 3/2\, C_4 [k_z (\varepsilon_{xx}-\varepsilon_{yy})]$, $I=\sqrt{3}/2\, C_4 [k_x(\varepsilon_{yy}-\varepsilon_{zz})+i k_y (\varepsilon_{xx}-\varepsilon_{zz})]$, and $J=1/\sqrt{3}\, D_u(\varepsilon_{xx}-\varepsilon_{yy})$. 
Here, $P$ is the matrix element of the inter-band momentum as defined in Ref.~\onlinecite{Winkler2003}, and $\varepsilon_{ii}$, $i=x,y,z$, are the diagonal components of the strain tensor. $D_d$ and $D_u$ denote deformation potentials and the constant $C_4$ is defined in Ref.~\onlinecite{Trebin1979}.
For simplicity, we restrict ourselves to a diagonal strain tensor $\bm{\varepsilon}$, since, due to  symmetry, the shear strain components are only of appreciable size at the dot interfaces and negligible everywhere else.
This assumption is valid because, due to their small effective mass, holes are strongly confined to the center of quantum dots \cite{Tadic2002}. 
%
%
In the vicinity of the $\Gamma$-point, the basis states in the single bands of the unperturbed Hamiltonian are given by
\begin{equation}
  |\Psi^{\pm}_{j,\mathbf{n}}\rangle = \phi^{\mathbf{n}}_j(\mathbf{r}) |\mathbf{u}^{\pm}_j(\mathbf{r}),\pm_j\rangle,
\end{equation}
where $j=\text{CB, HH, LH, SO}$ is the band index and $\pm$ distinguishes between the two states of each band which are degenerate in the bulk system. 
The basis functions of $H_{\text{K}}$ consist of $s$- and $p$-symmetric Bloch states $|\mathbf{u}^{\pm}_j(\mathbf{r})\rangle$ in the CB and VB, respectively, and spin states $|\pm_j\rangle$. 
The envelopes are given by the three-dimensional eigenfunctions of the harmonic confinement potential $V_{\text{conf}}$, $\phi^{\mathbf{n}}_j(\mathbf{r})$, with $\mathbf{n} = (n_x,n_y,n_z)$ being a vector of the according quantum numbers.
Motivated by the large energy splittings in quantum dots we choose $n_x, n_y, n_z\in\{0,1\}$.
%
%
We approximately block-diagonalize the complete Hamiltonian $H=H_{\text{K}}+V_{\text{conf}}+H_{\bm{\varepsilon}}$ in the HH subspace by a Schrieffer-Wolff transformation $\tilde{H} = e^{-A} H e^{A}$. 
The eigenstates of the diagonal HH subsystem are determined by
\begin{equation}
  |\Psi_{\text{hyb}}\rangle  = |\tilde{\Psi}\rangle \simeq \left(\mathds{1}-A^{(1)}\right)|\Psi\rangle,  
\end{equation}
where $A^{(1)}$ is the anti-hermitian, block off-diagonal matrix describing the Schrieffer-Wolff transformation to first order. 
Explicitly, the hybridized eigenstates of the effective 2$\times$2 HH Hamiltonian read
\begin{eqnarray}
  |\Psi_{\text{hyb}}^{\tau}(\bm{\varepsilon})\rangle &=& \mathcal{N} \sum_{\substack{j,\mathbf{n},\tau'}}\lambda^{\tau'\!,\tau}_{j, \mathbf{n}}(\bm{\varepsilon})  |\Psi^{\tau'}_{j,\mathbf{n}}\rangle,
\end{eqnarray}
$\tau,\tau'=\pm$, with $\lambda^{\tau',\tau}_{j, \mathbf{n}}(\bm{\varepsilon}) = \langle\Psi^{\tau'}_{j,\mathbf{n}}|H|\Psi^{\tau}_{\text{HH},\mathbf{0}}\rangle/(E_{j,\mathbf{n}}-E_{\text{HH},\mathbf{0}})$ being overlap matrix elements, where $H$ and $E_{j,\mathbf{n}}$ introduce the strain dependence. 
$E_{j,\mathbf{n}}$ is the eigenenergy of the state $|\Psi^{\tau'}_{j,\mathbf{n}}\rangle$ and $\mathcal{N}$ ensures proper normalization.
In the zero strain case we find for $|\Psi_{\text{hyb}}^{\tau}(0)\rangle$ the leading coefficients $|\lambda^{\tau,\tau}_{\text{CB},(0,1,0)}(0)|=|\lambda^{\tau,\tau}_{\text{CB},(1,0,0)}(0)|\simeq0.11$, $|\lambda^{\tau,\tau}_{\text{LH},(0,1,1)}(0)|=|\lambda^{\tau,\tau}_{\text{LH},(1,0,1)}(0)|\simeq0.097$, and $|\lambda^{\tau,\tau}_{\text{SO},(0,1,1)}(0)|=|\lambda^{\tau,\tau}_{\text{SO},(1,0,1)}(0)|\simeq0.031$. 
For all configurations, $\lambda^{\tau,\tau}_{\text{HH},\mathbf{0}}(\bm{\varepsilon})=1$.
The system parameters used in the calculations are listed in Table \ref{tab:ParamInAs}.
\begin{table}
 \begin{tabular}[b]{cdccd}
  \hline\hline
  $E_g$		& 0.418 \text{eV} \text{\cite{Winkler2003}}	& & $D_d$		& 1.0\text{eV} \text{\cite{Vurgaftman2001}}	\\
  $m'$		& 0.026 m_0	& & $D_u$		& 2.7\text{eV}\text{\cite{Winkler2003}}	\\
  $P$		& 9.197 \text{eV\AA}\text{\cite{Winkler2003}}	& & $D_u'$	& 3.18\text{eV} \text{\cite{Winkler2003}}	\\
  $\gamma_1$	& 20.0 \text{\cite{Vurgaftman2001}}	& & $C_4$		& 11.3\text{eV\AA} \text{\cite{Ranvaud1979}}	\\
  $\gamma_2$	& 8.5 \text{\cite{Vurgaftman2001}}	& & $C_5'$	& 18.4 \text{eV\AA} \text{\cite{Ranvaud1979}}\\
  $\gamma_3$	& 9.2 \text{\cite{Vurgaftman2001}}	& &&\\
  $\alpha$ 	& 0.666					& & $a_{\text{InAs}}$ & 6.058 \text{\AA}\text{\cite{Winkler2003}}\\
  \hline\hline 
  \end{tabular}
  \caption{Values of InAs parameters we use as input for the 8$\times$8 Hamiltonian $H_{\text{K}}+V_{\text{conf}} + H_{\bm{\varepsilon}}$.}
  \label{tab:ParamInAs}
\end{table}
%
%
\section{Effective Hyperfine Hamiltonian of the Heavy Hole spin\label{sec:hyperfine}}
The hybridized HH states couple to the $k$th nucleus by the Fermi contact interaction $h_1^k$, being non-negligible due to the $s$-symmetric CB admixtures, the anisotropic hyperfine interaction $h_2^k$, and the coupling of the orbital angular momentum (OAM) to the nuclear spins $h_3^k$ (see Refs.~\onlinecite{Stoneham1972,Fischer2008}).
We derive an effective, strain dependent hyperfine Hamiltonian in the HH subspace by taking matrix elements over a single Wigner-Seitz (WS) cell: $H_{\text{eff}}^{\tau, \tau'}(\bm{\varepsilon}) = \sum_k \langle\Psi_{\text{hyb}}^{\tau}(\bm{\varepsilon})|\sum_{i=1}^3 h_i^k|\Psi_{\text{hyb}}^{\tau'}(\bm{\varepsilon})\rangle_{\text{WS}}$, $\tau,\tau'=\pm$. 
For the numerical evaluation of the matrix elements we model the WS cell as a sphere of radius one half of the \mbox{In-In} atom distance, centered in the middle of the InAs bond. 
The basis functions of $H_{\text{K}}$, $|\mathbf{u}^{\pm}_j(\mathbf{r}),\pm_j\rangle$, are written as products of OAM eigenstates and spin states \cite{Winkler2003}.
We approximate the eigenstates of OAM, $S$, $P^z$, and $P^{\pm}$, as linear combinations of atomic eigenfunctions\cite{Gueron1964}, $\mathbf{u}^{\pm}_j(\mathbf{r}) = \alpha \psi_{\text{In}}^{5lm}(\mathbf{r}+\mathbf{d}/2)\pm \sqrt{1-\alpha^2}\psi_{\text{As}}^{4lm}(\mathbf{r}-\mathbf{d}/2)$, where $\alpha$ is the electron distribution between the two atoms and $\psi^{nlm}(\mathbf{r}) = R_{nl}(r)Y_l^m(\vartheta, \varphi)$ are hydrogenic eigenfunctions with quantum numbers $n$, $l$, and $m$.  
The radial part of the wavefunction depends on the effective central charge $Z_{\text{eff}}$ of the nuclei where we use values for free atoms \cite{Clementi1963, Clementi1967}. 
$\mathbf{r}\pm\mathbf{d}/2$ denotes the position of the hole with respect to the nuclei located at $\pm\mathbf{d}/2$ in the WS cell, where $\mathbf{d}=a_{\text{InAs}}(1,1,1)/4$ is the InAs bonding vector defined by the lattice constant $a_{\text{InAs}}$. 
The bonding and anti-bonding character of the VB and CB are expressed by the $+$ and $-$ signs, respectively, and $\int_{\mathrm{WS}} \mathrm{d}^3 r |\mathbf{u}_j^{\pm}(\mathbf{r}),\pm_j|^2=2$ enforces normalization \cite{Coish2009}.
The error of this method is small and has been estimated in Ref.~\onlinecite{Fischer2008}. 
For strain distributions in the vicinity of the realistic strain configuration of a cylindric InAs quantum dot, i.e.\ $\varepsilon_{xx}=\varepsilon_{yy}=-0.06$ and $\varepsilon_{zz}=0.06$ (see Ref.~\onlinecite{Tadic2002}), we find an effective hyperfine Hamiltonian of the form
\begin{eqnarray}
  H_{\text{eff}} &=& (b_z + h^z)S^z+\frac{1}{2}(h^+ S^- + h^- S^+). \label{eq:effHam}
\end{eqnarray}
Here, the term proportional to $b_z = g_h \mu_B B$ accounts for the Zeeman splitting due to a magnetic field $B$ along the growth direction, with $g_h \simeq 2$ being the HH g factor and $\mu_B$ the Bohr magneton. 
The components of the Overhauser field read $h^{z,\pm} = \sum_k A^{z,\pm}_k(\bm{\varepsilon}) I_k^{z,\pm}$, where $A^{z,\pm}(\bm{\varepsilon}) = \sum_i \nu_i A^{z,\pm}_i(\bm{\varepsilon})$ denote the corresponding strain dependent hyperfine coupling constants weighted by the nuclear abundance $\nu_i$ of each atomic species $i$. 
$\bm{S}$ is the pseudospin $1/2$ operator of the hybridized HH states and $\bm{I}_k$ is the nuclear spin operator of the $k$th nucleus. 
We find for the effective hyperfine coupling
\begin{eqnarray}
  A^z_k(\bm{\varepsilon}) &\simeq& v_0 A_{z}(\bm{\varepsilon}) \left|\phi_{\text{HH}}^{\mathbf{0}}(\mathbf{r}_k)\right|^2,\label{eq:Az}\\
  A^{\pm}_k(\bm{\varepsilon}) &\simeq& \sum_{\substack{j,j'\!,\mathbf{n},\mathbf{n}'}} v_0  A_{\pm,j,j'}(\bm{\varepsilon})\phi_{j}^{\mathbf{n}}(\mathbf{r}_k)^* \phi_{j'}^{\mathbf{n}'}(\mathbf{r}_k),\label{eq:Apm} 
\end{eqnarray}
where $v_0$ is the volume occupied by a single nucleus. 
$A_z(\bm{\varepsilon})$ and $A_{\pm,j,j'}(\bm{\varepsilon})$ are the hyperfine coupling strengths and are given by $\mathbf{A}_{j,j'}(\bm{\varepsilon})\cdot\mathbf{I}^k=\sum_{\kappa, \kappa'}(\lambda^{\kappa,\tau}_{j,\mathbf{n}}(\bm{\varepsilon}))^*\lambda^{\kappa',\tau'}_{j',\mathbf{n}'}(\bm{\varepsilon})\langle\mathbf{u}_j^{\kappa}(\mathbf{r}),\kappa_{j}|\sum_{i=1}^3 h_i^k|\mathbf{u}_{j'}^{\kappa'}(\mathbf{r}),\kappa'_{j'}\rangle$, where $\kappa,\kappa'=\pm$ and  $\mathbf{I}^k$ is the nuclear spin operator. 
In Eq.~(\ref{eq:Apm}), we neglect contributions where $A_{\pm,j,j'}(\bm{\varepsilon})$ is more than one order of magnitude smaller than the leading term. 
We find $|A_z(\bm{\varepsilon})|\gg \mathrm{max}|A_{\pm,j,j'}(\bm{\varepsilon})|$, thus $H_{\text{eff}}$ is predominantly of Ising form with additional small pair-flip processes between nuclear and hole spin. 
%
%
\section{Effect of strain on the Heavy Hole spin dynamics\label{sec:timeevolution}}
For a Hamiltonian of the form of $H_{\text{eff}}$, the time evolution of the $S^+(t)$ component and hence the decoherence of the HH pseudospin state is described by the Nakajima-Zwanzig master equation \cite{Coish2004}.
We obtain an algebraic form in the rotating frame with frequency $\omega_n$ by performing a Laplace transform, $f(s) = \int_0^{\infty}f(t)e^{-st}\mathrm{d}t,\,\mathrm{Re}[s]>0$, yielding
\begin{equation}
  S^+(s+i \omega_n) = \frac{\langle S^+\rangle_0}{s+\Sigma (s+i \omega_n)}.\label{eq:LaplaceNakajimZwan} 
\end{equation}
Here $\langle S^+\rangle_0=\mathrm{Tr}\,S^+ \rho$ with density operator  $\rho$ and $\Sigma(s)$ is the Laplace transformed memory kernel which describes the dynamics of $ S^+$ and is derived in Refs.~\onlinecite{Coish2004,Coish2010}. 
The Zeeman splitting $\omega_n$ is determined by the eigenvalue equation $\omega_n |n\rangle = (b_z+h^z)|n\rangle = (g_h \mu_B B + p A_z(\bm{\varepsilon}) I)|n\rangle$, where $p$ $(|p|\leq1)$ is the polarization of the nuclear spins in positive $z$ direction, and $|n\rangle$ is a narrowed nuclear spin state \cite{Coish2004}. 
The exact Eq.~(\ref{eq:LaplaceNakajimZwan}) can only be solved perturbatively by expanding $\Sigma(s)$ in powers of the flip-flop processes $V =(h^+ S^- + h^- S^+)/2$. 
This is possible since the  energy scale of $V$ is much smaller than the one associated with the Ising term $\sim h^z$ in Eq.~(\ref{eq:effHam}). 
Following Ref.~\onlinecite{Coish2010}, we expand $\Sigma(s)$ up to fourth order in $V$, $\Sigma(s) =  \Sigma^{(2)}(s)+\Sigma^{(4)}(s)+\mathcal{O}(V^6)$, where the Zeeman splitting between the HH and nuclear spins forbids processes of odd order. 
$\Sigma^{(2)}(s)$ and $\Sigma^{(4)}(s)$ are given explicitly in Eqs.~(\ref{eq:sigma2}) and (\ref{eq:sigma4}) of the Appendix.
$\Sigma^{(2)}(s)$ is purely real and hence leads to no decay in Eq.~(\ref{eq:LaplaceNakajimZwan}) but to a frequency shift $\Delta\omega = -\mathrm{Re}\left[\Sigma^{(2)}(s+i \omega_n)\right]$. 
This reflects the fact that energy conservation forbids the real flip of the electron spin, and only virtual flips are possible. 
The imaginary part of $\Sigma^{(4)}(s)$ yields a decay, resulting in the decoherence rate $1/T_2$ given by the relation $1/T_2 = -\mathrm{Im}\left[\Sigma^{(4)}(i \omega_n+i \Delta\omega-0^+)\right]$, where $0^+$ is a positive infinitesimal. 
$\Sigma^{(2)}(s)$ and $\Sigma^{(4)}(s)$ are evaluated in the continuum limit (see Appendix).
We simplify the calculations by averaging over the vertical dependence of the coupling constants $A_k^{z,\pm}(\bm{\varepsilon})$, which is possible since $a \ll L$.
The frequency shift $\Delta\omega$ can be calculated directly, whereas the lengthy calculation of the decoherence rate $1/T_2$ can be found in the Appendix. 
After a calculation analogously to Ref.~\onlinecite{Fischer2010} we find
\footnote{Here, we have corrected an error in the previous calculation\cite{Fischer2010}. As a consequence of this, the dip in Fig.~1 of Ref.~\onlinecite{Fischer2010} turns out to be an artifact. }
\begin{eqnarray}
   \frac{1}{T_2} &=& \pi\frac{ c_{+} c_{-} }{4 \omega_n^2} \frac{|A_{\pm}|^4}{|A_z|}  \int_{\eta}^1 \mathrm{d}x x [\ln x]^2 (x-\eta) [\ln (x-\eta)]^2, \nonumber\\ \label{eq:finalT2}
\end{eqnarray}
where $c_{\pm} = I(I+1)-\langle\langle m(m\pm1)\rangle\rangle$ with nuclear spin $I$ and $m=-I,\ldots,I$. The brackets $\langle\langle\ldots\rangle\rangle$ denote averaging over all eigenvalues $m$ of $I_k^z$.
$\eta(\bm{\varepsilon}) = \Delta\omega/|A_z| \propto1/\omega_n$ and $1/T_2$ can be evaluated numerically for any $\eta<1$.
It is evident that the Ising-like form of the Hamiltonian (\ref{eq:effHam}), which corresponds to $|A_z|\gg|A_{\pm}|$, prolongs $T_2$. 
\begin{figure}[t!]
  \centering
  \includegraphics[width=\columnwidth]{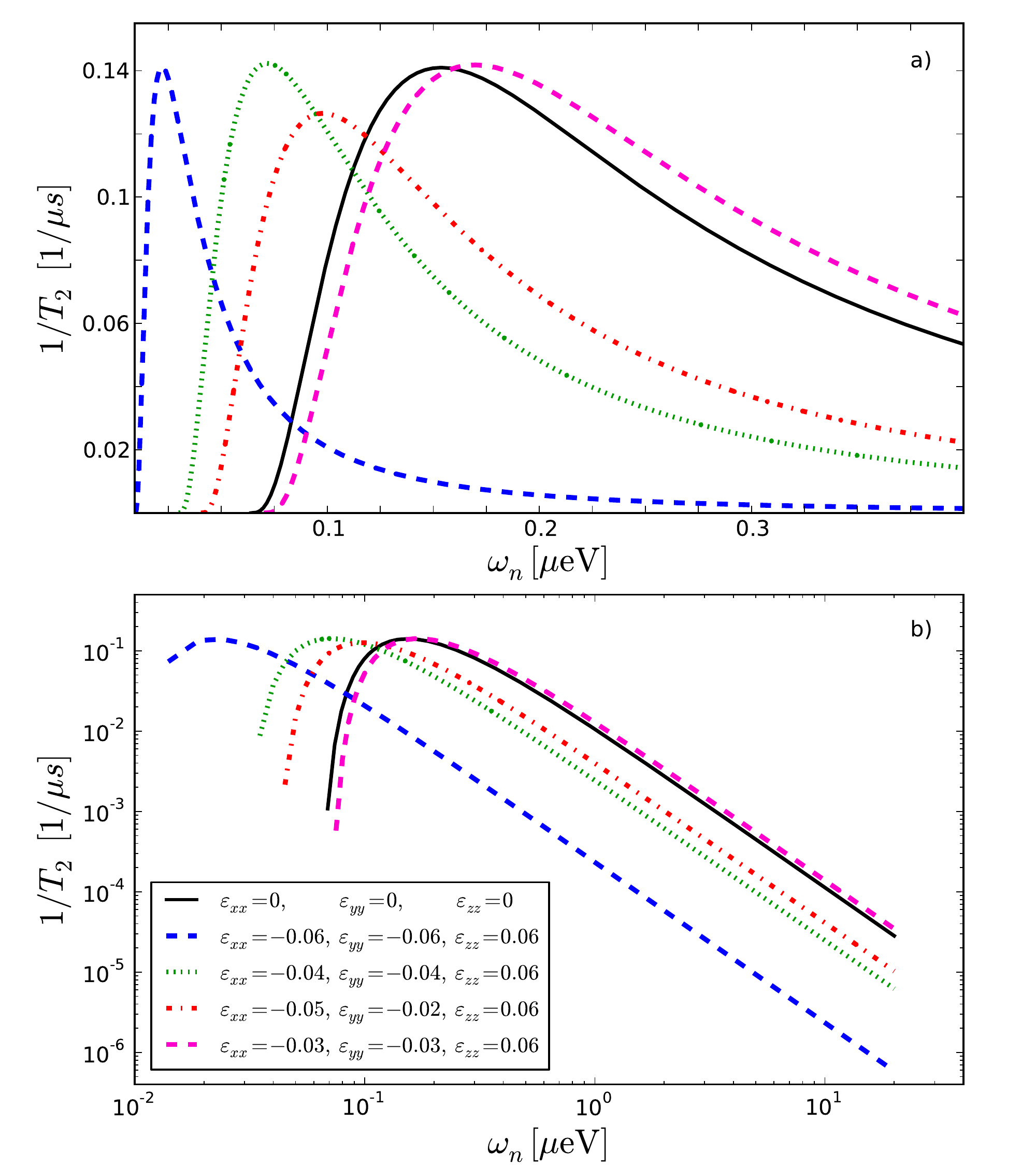}
  \caption{The decoherence rate $1/T_2$ for different internal strain configurations as a function of the Zeeman energy $\omega_n$. (a) For most configurations the maximum is shifted to the left with respect to the unstrained case except if $\mathrm{Tr}\,\bm{\varepsilon}\gtrsim\xi$. (b) Large $\omega_n$, i.e.\ magnetic fields $B$ or polarizations $p$, cause a power-law decay of $1/T_2$. The legend in the lower panel is valid for both plots and denotes the strain configurations. 
  \label{fig:DecohRates}}
\end{figure}
The effect of non-zero strain configurations on the hyperfine decoherence rate is clearly visible when comparing with the zero strain case.
In Fig.~\ref{fig:DecohRates} we dis\-play the decoherence rate $1/T_2$ as a function of the Zeeman energy $\omega_n = g_h \mu_B B + p A_z(\bm{\varepsilon}) I$ for an unstrained dot and different internal strain configurations.
The general shape of the decoherence rate remains unchanged for the different strain distributions, but the rate is shifted along the $\omega_n$-axis, the width of the peak is altered, and a lowering of the rate's maximum for asymmetric lateral strain is induced.
The lower bound of the rates on the $\omega_n$-axis is determined by $\eta = 1$. 
The decoherence rate is shifted to the left with respect to the zero-strain curve for $\mathrm{Tr}\,\bm{\varepsilon}<\xi$ and shifted to the right if $\mathrm{Tr}\,\bm{\varepsilon}\gtrsim\xi$, respectively, where $\xi$ is a small negative number of $\mathcal{O}(10^{-3})$. 
The latter relation corresponds to a dominant vertical strain tensor component.
When the rate is shifted to the left, the peak becomes more pronounced, hence the sensitivity of $1/T_2$ to changes in the external parameters is increased.
The associated hybridized wavefunctions show a gradual lowering of the admixtures of all leading components with respect to the zero-strain case. 
A broadening of the peak occurs when the curve is shifted to the right. 
Here, the CB admixture of the hybridized wavefunction is increased while the other admixtures are lowered again. 
\begin{figure}[t]
  \centering
  \includegraphics[width=\columnwidth]{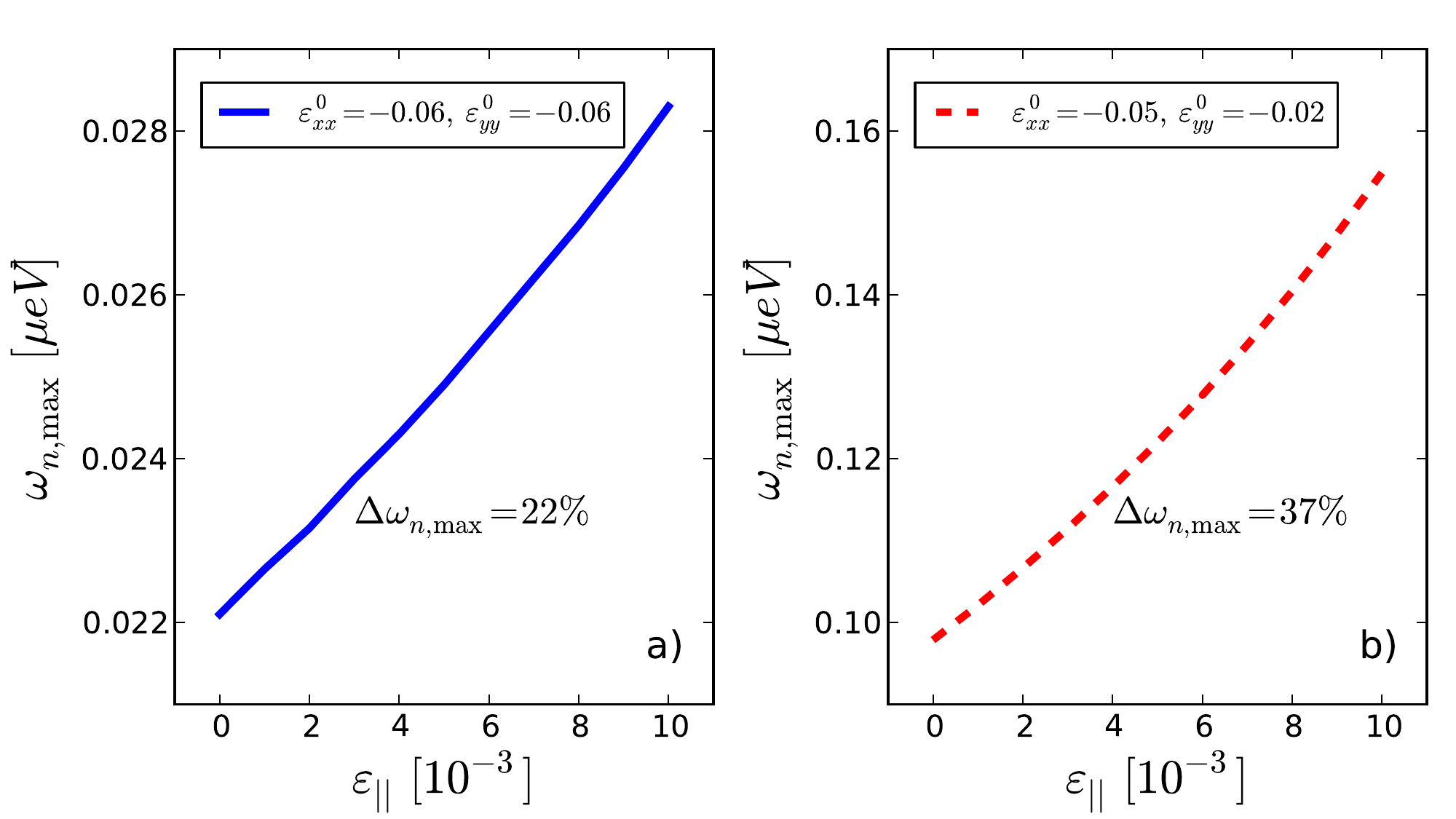}
  \caption{Shift of the peak of the decoherence rate located at $\omega_{n,\text{max}}$ when external strain $\varepsilon_{\|}$ up to the breaking limit is applied. In both panels the change is roughly linear. For the asymmetric lateral strain configuration (b) the relative change of $\omega_{n,\text{max}}$ is considerable larger than for the symmetric configuration (a). The legends display the internal strain configurations of the dots, $\varepsilon_{xx}^0$ and $\varepsilon_{yy}^0$, at $\varepsilon_{\|}=0$. For all configurations we keep $\varepsilon_{zz}=0.06$ constant. 
  \label{fig:DecohRate_shiftPiezo}}
\end{figure}
For the chosen dot geometry, $L=10\text{nm}$ and $a=2\text{nm}$, the minimal coherence time at the peak of the curves is  $T_2\simeq 7 \mu\text{s}$. 
For large magnetic fields $B$ or polarizations $p$, the curves decay following a power-law as evident in  Fig.~\ref{fig:DecohRates}b). 
As a general result, we state that, regardless the strain configuration, the decoherence rate $1/T_2$ can be decreased over orders of magnitude by relatively small changes of the external parameters. \\
%
%
%
The strain fields in a quantum dot can be modified by applying additional strain, e.g.\ by the technique demonstrated in Ref.~\onlinecite{Seidl2006}. 
Here, a GaAs sample containing InAs quantum dots is tightly glued on top of a piezoelectric stack, its stretching direction aligned with the $\left<110\right>$ crystal axis, and a voltage is applied. 
So far, additional strain of about $\varepsilon_{\|}\simeq0.003$ (see Ref.~\onlinecite{Ding2010}) has been reached whereas the breaking point of GaAs corresponds to a strain of $\varepsilon_{\|}\approx0.012$ (see Ref.~\onlinecite{Bhargava1967}).
We examine the peak of the decoherence rate located at $\omega_{n,\text{max}}$ which is determined by the implicit, strain dependent equation
\begin{equation}
\int_{\eta}^1\mathrm{d}x x [\ln x]^2 \ln(x-\eta)\left[\left(\frac{3}{2}\eta-x\right)\ln(x-\eta)+\eta\right] =0.  
\end{equation}
Additional strain alters $\omega_{n,\text{max}}$ significantly, as displayed in Fig.~\ref{fig:DecohRate_shiftPiezo}, and hence inflicts measurable changes on the decoherence rate.
The relative shift of the peak for the asymmetric lateral strain configuration is about 37\%, thus larger than for the symmetric configuration where the shift is 22\%. 
The strain-induced change of the decoherence rate is directly connected to variances in the magnitude of the coupling strengths $A_z(\bm{\varepsilon}) $ and $A_{\pm,j,j'}(\bm{\varepsilon})$. 
In Fig.~\ref{fig:Coupling_shiftPiezo} we display the dependence of the absolute values of the dominant coupling strengths on applied strain $\varepsilon_{\|}$.
We find that the relative change of $A_{\pm,\text{CB}}$, ranging up to 21.5\%, is the largest of all whereas the relative change of $A_{\pm,\text{LH}}$ is only about 2\%. 
The Ising-like coupling $A_{z}$ (not on display) changes less than 1\% and thus is negligible. 
From this we deduce that the strain-induced changes of the decoherence rate can be attributed mainly to the difference in the CB admixture of the hybridized HH states. 
Hence the usually neglected contribution of the CB to the hole spin dephasing is of significance.
\begin{figure}[t]
  \centering
  \includegraphics[width=\columnwidth]{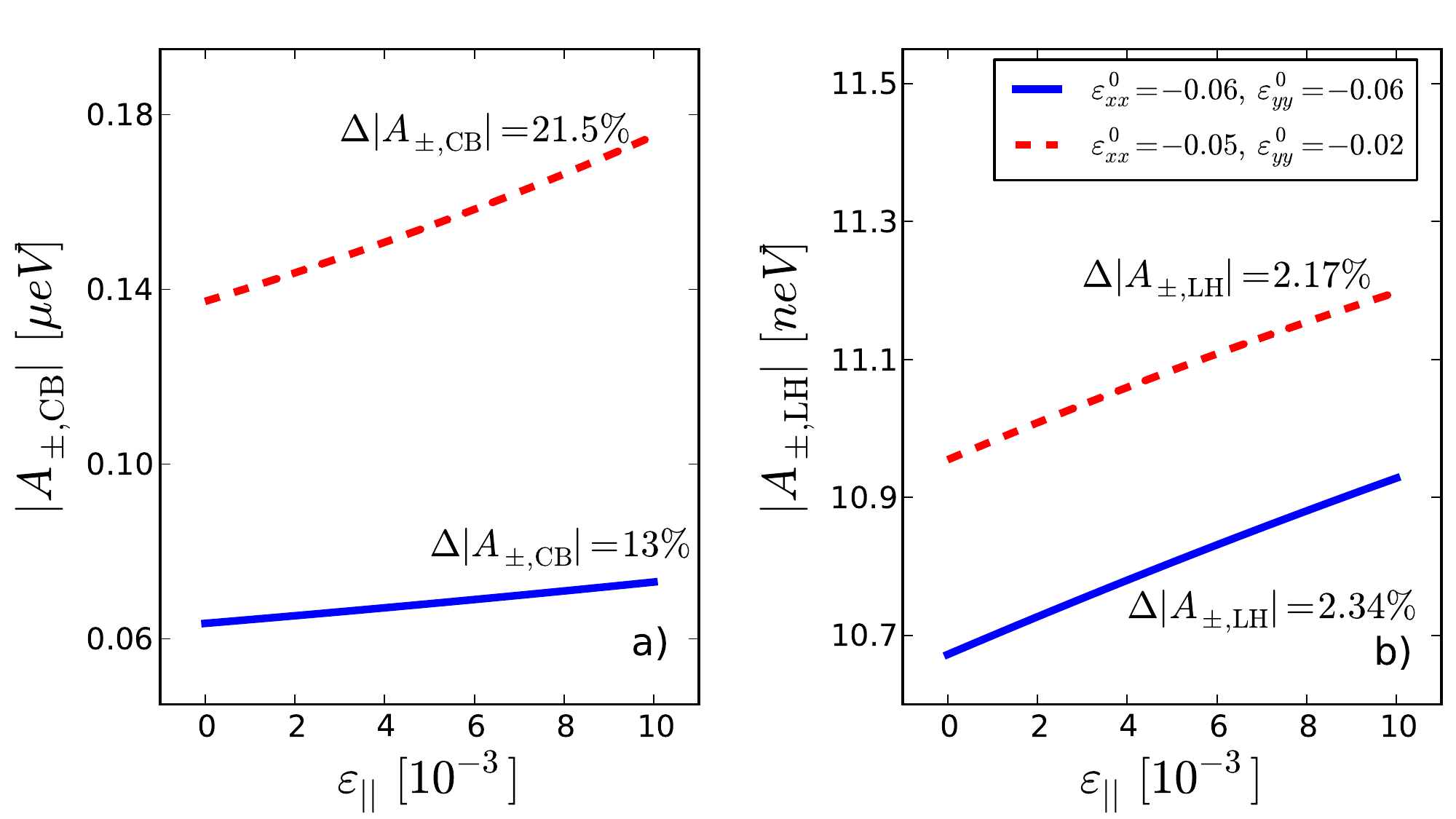}
  \caption{The largest contributions to the hyperfine coupling as a function of applied external strain $\varepsilon_{\|}$. The relative change of the CB coupling (a) is much larger than the change of the LH coupling (b). In the legend the internal strain configurations of the dots, $\varepsilon_{xx}^0$ and $\varepsilon_{yy}^0$, at $\varepsilon_{\|}=0$ are displayed. Again we keep $\varepsilon_{zz}=0.06$ constant for all configurations. 
  \label{fig:Coupling_shiftPiezo}}
\end{figure}
\section{Summary\label{sec:summary}}
In conclusion, we investigated the dynamics of hybridized HH spin states confined to self-assembled and hence strained semiconductor quantum dots. 
By taking into account hyperfine interaction between these states and the surrounding nuclei an effective, strain dependent Hamiltonian was found, which is, for realistic strain configurations, predominantly of Ising form.
The time evolution of its $S^+$ component was derived for a narrowed nuclear spin state, and we have shown that the internal strain fields of self-assembled quantum dots affect the decoherence rate $1/T_2$ significantly.  
For all strain configurations, $1/T_2$ was found to be tunable over orders of magnitude by adjusting external parameters.  
Different strain fields were shown to cause a shift of the dependence of $1/T_2$ on external parameters $\omega_n$ and to change the non-monotonicity of the peak. 
Additional application of external strain inflicted measurable changes upon $1/T_2$ which could mostly be  attributed to large alterations in the effective hyperfine coupling of the CB admixture.
This finding indicated the importance of the CB admixture of the hybridized HH states regarding the sensitivity of $1/T_2$ on strain. 
\begin{acknowledgments}
We thank Dimitrije Stepanenko and Richard J. Warburton for helpful discussions. This work has been supported by the Swiss SNF, NCCR Nanoscience, NCCR QSIT, and DARPA. 
\end{acknowledgments}
%

%
\FloatBarrier
\appendix
\section{Hamiltonian - Explicit form}
\FloatBarrier
In this work, we use the 8$\times$8 Kane Hamiltonian as given in Ref.~\onlinecite{Winkler2003}, Appendix C, and a harmonic confinement potential. 
For completeness, we display here the general structure and relevant parts. 
The Kane Hamiltonian is given by
\begin{equation}
  H_{\text{K}} = \left(\begin{array}{cccc}
    K_{11}	& K_1 & K_2 & K_3\\
    K_1^{\dagger}&K_{22}&K_4 & K_5\\
    K_2^{\dagger}&K_4^{\dagger}&K_{33}&K_6\\
    K_3^{\dagger}&K_5^{\dagger}&K_6^{\dagger}&K_{44}
  \end{array}\right),
\end{equation}
of which the blocks
\begin{equation}
\begin{array}{ll}
  K_{22} = \begin{pmatrix} A&0\\0&A\end{pmatrix}, &
 K_4 = \sqrt{3}\begin{pmatrix}2 C&D\\D^*&-2C^*\end{pmatrix},\\
&\\[-3mm]
   K_1^{\dagger} =  \begin{pmatrix} -B&0\\0&B^*\end{pmatrix},&
  K_5 = \sqrt{6}\begin{pmatrix}-C&-D\\D^*&-C^*\end{pmatrix},
\end{array}
\end{equation}
are relevant for the calculations.
The single entries are denoted by $A = -\hbar^2/(2 m_0)[(\gamma_1+\gamma_2)(k_x^2+k_y^2)  +(\gamma_1-2 \gamma_2) k_z^2]$, $B = 1/\sqrt{2}\, P (k_x-i k_y)$, $C = \hbar^2/(2 m_0) \gamma_3 k_z (k_x-i k_y)$, and $D = \hbar^2/(2 m_0)[\gamma_2 (k_x^2-k_y^2)-2 i \gamma_3 k_x  k_y]$, with $\hbar$ being Planck's constant, $m_0$ being the bare electron mass and $\gamma_i$, $i=1,2,3$, denoting the Luttinger parameters. 
In $ H_{\text{K}}$, terms proportional to $C_k$, $B_{7v}$ and $B_{8v}^{\pm}$ were omitted due to their smallness\cite{Fischer2008, Fischer2010}. 
The harmonic confinement potential 
\begin{equation}
 V_{\text{conf},j}(\mathbf{r}) = \left(\frac{m_{j,\perp} \omega_{j,\perp}^2}{2} z^2 + \frac{m_{j,\|} \omega_{j,\|}^2}{2} (x^2+y^2)\right) \, \mathds{1}_{2\times2}
\end{equation}
is defined by the confinement lengths $L$ and $a$ via $\omega_{j, \perp}=\hbar/(m_{j, \perp}a^2)$ and $\omega_{j, \|}=\hbar/(m_{j, \|}L^2)$. The effective masses in the single bands are given by $m_{\text{CB}, \perp/\|} = m',\, m_{\text{HH/LH}, \perp}= m_0/(\gamma_1 \mp 2 \gamma_2), m_{\text{HH/LH}, \|}= m_0/(\gamma_1 \pm \gamma_2)$ and $m_{\text{SO}, \perp/\|} = m_0/\gamma_1$. $\mathds{1}_{2\times2}$ denotes the 2$\times$2 unit matrix.
Table \ref{tab:ParamInAs} in the main text lists the material parameters of InAs we use in the calculations as input for the 8$\times$8 Hamiltonian $H_{\text{K}} + V_{\text{conf}} + H_{\bm{\varepsilon}}$. 
\section{Continuum limit of the memory kernel\label{sec:contlim}}
For the second and forth order in a homonuclear system rotating with $\omega_n$ we find \cite{Coish2010}, 
\begin{eqnarray}
  \Sigma^{(2)}(s+i \omega_n) &\simeq& -\frac{c_{+} + c_{-}}{4 \omega_n} \sum_{k} |A_k^{\pm}|^2, \label{eq:sigma2}\\
  \Sigma^{(4)}(s+i \omega_n) &\simeq& -i \frac{c_{+} c_{-}}{4 \omega_n^2} \sum_{k_1, k_2} \frac{|A_{k_1}^{\pm}|^2 |A_{k_2}^{\pm}|^2}{s+i (A_{k_1}^z-A_{k_2}^z)}\label{eq:sigma4}
\end{eqnarray}
where we dropped the strain dependence of $A_k^{\pm,z} = A_k^{\pm,z}(\bm{\varepsilon})$ for readability. 
In both equations, $c_{\pm} = I(I+1)-\langle\langle m(m\pm1)\rangle\rangle$, where $I$ is the nuclear spin, $m=-I,\ldots,I$ and the brackets $\langle\langle\ldots\rangle\rangle$ indicate averaging over all eigenvalues $m$ of $I_k^z$.
By taking the continuum limit $v_0 \sum_k = \int \mathrm{d}^3 r$ we replace the sums by integrals. 
The strain dependent frequency shift $\Delta\omega  = -\mathrm{Re}\left[\Sigma^{(2)}(s+i \omega_n)\right] \sim 10^{-18} \mathrm{eV}^2/\omega_n$ can be evaluated for Eqs.~(\ref{eq:Az}) and (\ref{eq:Apm}) in a straightforward fashion.
A recalculation of the result of Ref.~\onlinecite{Fischer2010} gives exact shape of $1/T_2$ after some transformations of its original form.
Starting from
\begin{equation}
  \Sigma^{(4)}(s+i \omega_n) \simeq- i \frac{c_{+} c_{-}}{4 \omega_n^2} \sum_{k_1, k_2} \frac{|A_{k_1}^{\pm}|^2 |A_{k_2}^{\pm}|^2}{s +i (A_{k_1}^z-A_{k_2}^z)},
\end{equation}
we first simplify by inserting the $z$ averaged coupling, justified by $L\gg a$ and then performing the two-dimensional continuum limit.  
In the resulting two-dimensional equation we shift to polar coordinates $x_j = r_j \cos \varphi_j$ and $y_j = r_j \sin\varphi_j$.
After the angular integration we rescale the radial variables by replacing $r_j = r_{jj} L$. In the resulting integral we substitute $e^{-r_{11}^2} = x$,  $r_{11}^4 = (\ln x)^2$, $e^{-r_{22}^2} = y$ and $r_{22}^4 = (\ln y)^2$ and take only the leading term:
\begin{eqnarray}
   \Sigma^{(4)}(s+i \omega_n)   &\simeq&  -i\frac{c_{+} c_{-} |A_{\pm}|^4}{4 \omega_n^2 A_z}  \int_0^1 \int_0^1 \mathrm{d}x\mathrm{d}y\nonumber\\
  &&\frac{ (\ln x)^2(\ln y)^2 xy}{ s/A_z+i  ( x- y)}, 
\end{eqnarray}
where $A_z = A_z(\bm{\varepsilon})$ and $A_{\pm}$ denotes the sum of all $A_{\pm,j,j'}(\bm{\varepsilon})$ contributing to the leading term. 
To calculate the decoherence rate, we have to take into account that $1/T_2=-\mathrm{Im}\left[\Sigma(i \omega_n+i \Delta\omega- 0^+)\right]$ and hence consider
\begin{eqnarray}
   \Sigma^{(4)}(i \omega_n&+&i \Delta\omega- 0^+)\nonumber\\
 &\simeq& -i\frac{c_{+} c_{-} |A_{\pm}|^4}{4 \omega_n^2 A_z}  \int_0^1 \int_0^1 \mathrm{d}x\mathrm{d}y\nonumber\\
  &&\frac{ (\ln x)^2(\ln y)^2 xy}{ (i \Delta\omega -0^+)/A_z+i  ( x- y)}.
\end{eqnarray}
Since we are only interested in the imaginary part of the equation, we use the following relation
\begin{equation}
  \lim_{\chi\rightarrow 0}\frac{1}{\zeta \pm i \chi} = \mathcal{P}\frac{1}{\zeta}\mp i \pi \delta(\zeta),
\end{equation}
where $\chi,\zeta$ are real numbers and $\mathcal{P}$ indicates that in any following integration of the above expression the principle value of the integral has to be taken. 
We find
\begin{eqnarray}
   \mathrm{Im}\left[\right.&\Sigma^{(4)}&(i \omega_n+\left.i \Delta\omega- 0^+)\right]\nonumber\\
   &\simeq&  -\frac{c_{+} c_{-} |A_{\pm}|^4}{4 \omega_n^2 A_z} \int_0^1 \int_0^1 \mathrm{d}x\mathrm{d}y  (\ln x)^2 x\nonumber\\
  &&(\ln y)^2 y (-\pi)\delta(x-y+ \Delta\omega/A_z),
\end{eqnarray}
from which follows 
\begin{eqnarray}
  \frac{1}{T_2} &=& -\mathrm{Im}\left[\Sigma^{(4)}(i \omega_n+i \Delta\omega-	 0^+)\right]\\
  &=& \pi \frac{c_{+} c_{-}}{4 \omega_n^2} \frac{|A_{\pm}|^4}{|A_z|}  \int_0^1 \int_0^1 \mathrm{d}x\mathrm{d}y x y\nonumber\\
  && (\ln x)^2(\ln y)^2  \delta(x-y- \Delta\omega/|A_z|)\\
  &=& \pi \frac{c_{+} c_{-}}{4 \omega_n^2} \frac{|A_{\pm}|^4}{|A_z|}  \int_0^1 \mathrm{d}x  \left[\ln x\right]^2 \left[\ln (x-\eta)\right]^2\nonumber\\
  && x (x-\eta)\Theta(x-\eta)\Theta(1-x-\eta),
\end{eqnarray}
where we use $A_z =-|A_z| $ and replace $\eta = \Delta\omega/|A_z|$ in the final step. 
We arrive at the following expression for the decoherence time
\begin{eqnarray}
   \frac{1}{T_2} &=&\pi \frac{c_{+} c_{-}}{4 \omega_n^2} \frac{|A_{\pm}|^4}{|A_z|}  \int_{\eta}^1 \mathrm{d}x x [\ln x]^2 (x-\eta) [\ln (x-\eta)]^2,\nonumber\\ 
\end{eqnarray}
which can be evaluated numerically for any $\eta<1$.
%
%
%

\bibliographystyle{apsrev}

\end{document}